\documentclass[twocolappendix]{emulateapj}

\usepackage{amsmath,amssymb,amstext}
\usepackage{mathtools}
\usepackage{gensymb}
\usepackage[usenames,dvipsnames]{color}
\usepackage{tabularx}

\newcolumntype{Y}{>{\centering\arraybackslash}X}
\usepackage{threeparttable}
\usepackage{ulem}
\usepackage[colorlinks,urlcolor=blue,citecolor=blue,linkcolor=blue,pdfusetitle]{hyperref}

\usepackage{comment}
\usepackage{todonotes}
\shorttitle{Empirical Gap Formula}
\shortauthors{Duffell}

\begin{document}

\title{An empirically-derived formula for the shape of planet-induced gaps in protoplanetary disks}

\author{Paul C. Duffell}
\affil{Center for Astrophysics, Harvard \& Smithsonian, Cambridge, MA 02138}
\email{paul.duffell@cfa.harvard.edu}

\begin{abstract}
This study uses numerical hydrodynamics calculations and a novel method for densely sampling parameter space to measure the precise shape of a gap opened by a planet in a gaseous disk, as a function of planet-to-star mass ratio, disk Mach number, and disk viscosity.  Formulas for gap depth and width are determined, which are combined to form a complete formula for surface density as a function of radius in the disk.  This new analytical formula is compared with numerically-derived gaps opened by planets ranging from very low masses up to a few times Jupiter's mass, and excellent agreement is found over a wide range of parameter space relevant to planet-disk interactions.  A simple-to-use code is presented to rapidly generate synthetic disk profiles.

\end{abstract}


\section{Introduction} \label{sec:intro}

When a sufficiently massive planet is embedded in a gaseous disk, that planet carves out a low-density annulus in the disk, also known as a ``gap".  Gaps are important for a number of reasons; they might change the torques and accretion experienced by the planet \citep{1997Icar..126..261W, 2018ApJ...861..140K}, but more importantly, gaps can be directly observed.  In particular, the Atacama Large Millimeter Array (ALMA) has generated very detailed, resolved images of protoplanetary disks with distinct gap-like features \citep{2018ApJ...852..122H, 2018ApJ...869L..42H, 2018ApJ...869L..43H, 2018ApJ...869L..48G, 2018ApJ...869L..50P}.  This provides a potential means of discovering exoplanets during the formation process.

This motivates a complete theory of gap-opening by planets; that is, given a planet mass, local disk temperature, and viscosity, what is the surface density of the gas as a function of radius?

One can address this question numerically, for a single set of disk and planet parameters \citep[e.g.][]{2013ApJ...769...41D, 2014ApJ...782...88F}, but much more useful would be a simple 1D analytical model, so that one can easily find best-fit disk and planet parameters for a given disk observation.

The idea of constructing a 1D gap model is not new; the first such models were constructed assuming torque is deposited in the disk at the same place it is excited \citep{2004ApJ...612.1152V, 2006Icar..181..587C}.  These models did not always agree with numerically calculated gap profiles, especially in the ``shallow gap" regime.  In particular, they predicted that the planet needed to be above the thermal mass in order for it to open a gap, even in an inviscid disk.  Additionally, they predicted an exponential scaling of gap depth with planet mass, which is also not seen in numerical studies.

\cite{2012ApJ...755....7D} showed that low-mass planets can open gaps in sufficiently low-viscosity disks, and later found an empirical scaling for the gap depth as a function of planet mass, disk viscosity, and Mach number \citep{2013ApJ...769...41D}.  \cite{2014ApJ...782...88F} confirmed this result, and presented a ``zero-dimensional" interpretation for the scaling of \cite{2013ApJ...769...41D}.  \cite{2015MNRAS.448..994K} presented a disk model intended to improve upon past methods and agree with the scaling of \cite{2013ApJ...769...41D}.

\cite{2015ApJ...807L..11D} built a simple analytical disk model assuming all torque was deposited by shocks, using the semi-analytical scalings found by \cite{2001ApJ...552..793G} and \cite{2002ApJ...569..997R} to determine where the shock forms and how much angular momentum it deposits at different locations in the disk.  The gap model of \cite{2015ApJ...807L..11D} exhibited an unprecedented level of agreement between the model and numerical calculations for shallow gaps, but unfortunately the model was inaccurate for deep gaps.

Subsequently, disk models by \cite{2017PASJ...69...97K} attempted to fix this model in the deep gap regime, also incorporating an empirical result for the scaling of gap widths \citep{2015ApJ...806L..15K}.  \cite{2018MNRAS.479.1986G} calculated analytical gap profiles for wider gaps, taking into account that torque can be excited at the gap walls.

In this work, we present an approach similarly empirical in nature to \cite{2017PASJ...69...97K}, but using a novel method to derive more information from the numerical calculations than any previous disk-planet study.  Rather than keeping the planet mass fixed, or growing it self-consistently with some accretion prescription, the planet-to-star mass ratio is very slowly increased by hand from $q = 10^{-6}$ to a few times $10^{-3}$.  By increasing the mass ratio adiabatically, the planet always experiences an effectively steady-state disk, and when measuring the gap properties as a function of time, one effectively measures the gap properties as a continuous function of mass ratio.  In other words, parameter space is densely sampled for all values of $q$ in this range.

The new calculations make it possible to continuously sample gap depth and width as a function of mass ratio, greatly increasing the ability to extract analytical scalings.  From this, a new gap model is constructed, which agrees with numerical calculations for all planet masses up through a few times Jupiter's mass.

The numerical method will be presented in Section \ref{sec:numerics}, including an explanation of the diagnostics measured from the numerical calculations.  Analysis of gap depths and widths will be presented in Section \ref{sec:anly}, and a 1D gap model will be constructed from this analysis.  Finally, in Section \ref{sec:compare} this analytical gap model will be compared with steady-state numerical calculations from previous studies.  In Section \ref{sec:disc} the results will be discussed, and future improvements suggested.  Additionally, a simple-to-use code will be presented for rapidly generating synthetic disk profiles.

\section{Methods} \label{sec:numerics}

This study uses the moving-mesh code \texttt{Disco} \citep{2016ApJS..226....2D} to numerically integrate the equations of isothermal hydrodynamics in 2D:

\begin{equation}
\partial_t \Sigma + \nabla \cdot (\Sigma \vec v) = 0
\end{equation}
\begin{equation}
\partial_t ( \Sigma v_j ) + \nabla \cdot ( \Sigma \vec v v_j + P \hat x_j - \nu \Sigma \vec \nabla v_j ) = - \Sigma \vec \nabla \phi
\end{equation}
\begin{equation}
P = c^2 \Sigma
\end{equation}
where $\Sigma$ is surface density, $P$ is pressure, $\vec v$ is velocity, $\nu$ is the kinematic viscosity, $c$ is the local sound speed, and $\phi$ is the gravitational potential from the planet and central star.

The numerical method and treatment of planetary potential is the same as described in previous studies using \texttt{Disco} \citep[e.g.][]{2012ApJ...755....7D, 2015ApJ...806..182D}.  An important difference in the present study is that a novel approach is employed for the planetary potential: rather than choosing a single mass ratio and running the calculation until the disk is in steady-state, the mass ratio is very slowly increased from $q = 10^{-6}$ to $q = 2 \times 10^{-3}$, in a sufficiently slow, adiabatic fashion such that at any single time the planet can be considered to be in a steady-state disk:

\begin{equation}
    q(t) = q_0 e^{t/\tau},
\end{equation}
with $q_0 = 10^{-6}$ and $\tau = 10^4/\Omega$.

The benefit of this is that one can measure the gap depth and shape continuously as a function of mass ratio, rather than performing many separate numerical calculations to sample a discrete number of points.  Measuring a continuous curve rather than discretely sampled points can be essential for measuring scalings, especially ``break points" where the scalings change.

A happy by-product of this method is that it is much more well-behaved in a numerical sense; it has been previously shown that introducing the planetary potential too quickly can artificially excite vortices in the disk \citep{2017MNRAS.466.3533H}.

The main diagnostics being measured are the gap depth, and the inner and outer radii at several different depths.  The gap depth $\Sigma_{\rm gap}$ is simply the azimuthally-averaged surface density at the bottom of the gap.  The contribution to this density due to high-density material very close to the planet is excised by only considering fluid elements in the half of the disk opposite the planet when computing this azimuthal average.  The inner and outer radii of the gap $R_{\pm}^{0.9}$, $R_{\pm}^{0.5}$, and $R_{\pm}^{0.1}$ are also measured, defined by

\begin{equation}
\Sigma( R_{+}^{d} ) = \Sigma( R_{-}^{d} ) = d \Sigma_0.
\end{equation}

This will make it possible to measure the complete shape of the gap, using the position of the edges of the gap at different thresholds, for different values of $d$.

\section{Analysis of Numerical Results}
\label{sec:anly}

\subsection{Gap Depth}

\begin{figure}
\includegraphics[width=3.4 in]{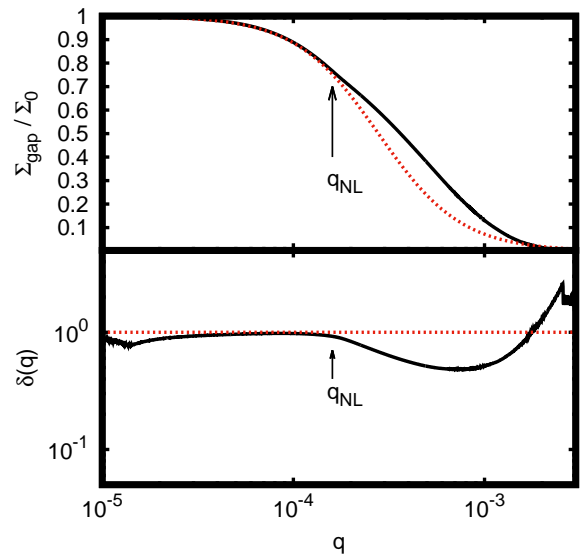}
\caption{Top Panel: Gap Depth as a function of mass ratio for the fiducial disk, $\mathcal{M} = 20$, $\alpha = 0.01$.  The red dashed curve is the \cite{2013ApJ...769...41D} scaling.  This begins to deviate at the mass ratio $q_{\rm NL}$.  Bottom Panel: This deviation is more clearly shown by plotting $\delta(q)$ (equation \ref{eqn:delta_def}).}
\label{fig:gapdepth1}
\end{figure}

\begin{figure}
\includegraphics[width=3.4 in]{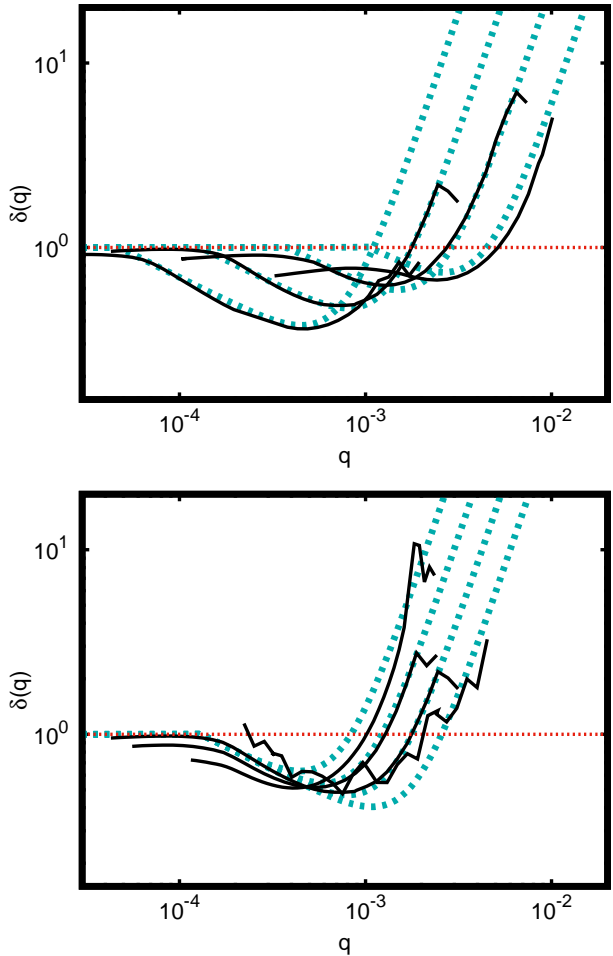}
\caption{$\delta(q)$ is plotted for a range of values of $\mathcal{M}$ and $\alpha$.  Best-fit analytical scalings for $\delta(q)$ are shown in the dashed curves.  For clarity in presentation, numerical curves in this plot are cut off just after the disk becomes unstable.  Effects on the gap depth due to instability are shown in Figure \ref{fig:gapdepth3}.}
\label{fig:gapdepth2}
\end{figure}

\begin{figure}
\includegraphics[width=3.4 in]{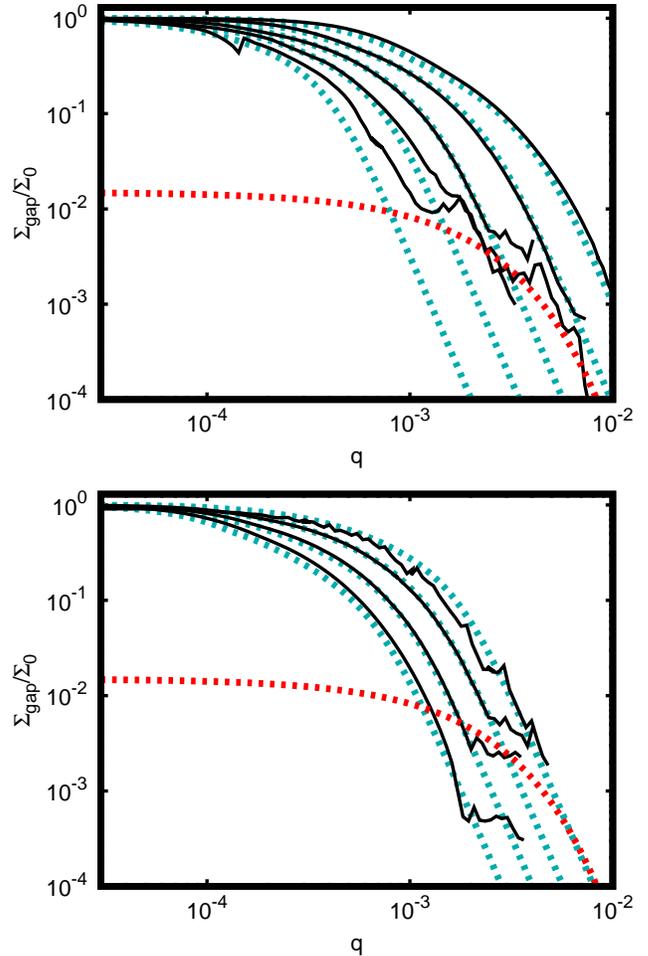}
\caption{Equation \ref{eqn:depth} for the gap depth is plotted using the best-fit formula for $\delta(q)$, compared with the numerical gap depth found for a range of disk-planet models.  The red dashed curve here represents a possible ``floor" which many of the models appear to hit when the disk becomes unstable.  The unstable regime is outside the scope of this study.}
\label{fig:gapdepth3}
\end{figure}

Let us first consider gap depth as a function of mass ratio.  \cite{2013ApJ...769...41D} provided an empirical formula which was later explained via the analytical gap models of \cite{2015MNRAS.448..994K} and \cite{2015ApJ...807L..11D}:

\begin{equation}
\Sigma^{\rm DM} / \Sigma_0 = \frac{1}{ 1 + \frac{0.45}{3 \pi} \frac{ q^2 \mathcal{M}^5 }{\alpha} }.
\end{equation}

This scaling applies to low-mass planets, whose planetary torques can be reasonably approximated in linear theory.  Deviations from the scaling of \cite{2013ApJ...769...41D} begin at mass ratio of a few times $10^{-4}$ (See Figure \ref{fig:gapdepth1}).  This can be most easily demonstrated by defining the function $\delta(q)$ such that

\begin{equation}
\Sigma / \Sigma_0 = \frac{1}{ 1 + \frac{0.45}{3 \pi} \frac{ q^2 \mathcal{M}^5 }{\alpha} \delta(q) }.
\label{eqn:depth}
\end{equation}

In other words, $\delta(q)$ is defined as

\begin{equation}
\delta(q) \equiv \frac{ \Sigma_0/\Sigma - 1 }{ 0.45 q^2 \mathcal{M}^5 / ( 3 \pi \alpha)}
\label{eqn:delta_def}
\end{equation}

The reason for invoking this particular expression is that $\delta(q)$ should be unity as long as the \cite{2013ApJ...769...41D} scaling applies, so one can simply measure the break point where it deviates from unity, and find the new scaling after that point.  This is shown in Figure \ref{fig:gapdepth1}, and again in Figure \ref{fig:gapdepth2} for several choices of disk mass ratio and viscosity, to demonstrate the scaling of the break point with $\mathcal{M}$ and $\alpha$.

If one can develop an empirical formula for $\delta(q)$, then one immediately has a complete formula for gap depth as a function of disk and planet parameters.  Figure \ref{fig:gapdepth1} shows the deviation of $\Sigma_{\rm gap}$ from the scaling of \cite{2013ApJ...769...41D}.  This deviation occurs at a particular mass ratio which will be called ``$q_{\rm NL}$" as it will be shown to coincide with the threshold for strong nonlinearity in the spiral wave.  

The lower panel of Figure \ref{fig:gapdepth1} plots $\delta(q)$, which shows this deviation more clearly.  It is possible to fit $\delta(q)$ with some power-laws; this gives the function

\begin{equation}
    \delta(q) =  \left\{ \begin{array}
			{l@{\quad \quad}l}
			1 & q < q_{\rm NL}\\  
    		(q/q_{\rm NL})^{-1/2} + (q/q_w)^3 & q > q_{\rm NL} \\ 
	\end{array} \right. 
	\label{eqn:delta}
\end{equation}

To determine $\delta(q)$ generally, we must first see how this scaling changes when $\mathcal{M}$ and $\alpha$ are varied.

Note that when $q$ becomes sufficiently large, the $\Sigma_{\rm gap}(q)$ scaling changes again, but the curve becomes much more erratic.  At this point, instabilities have been excited in the disk by the planet.  This will not be taken into account in our formula for $\delta(q)$ but a correction will be suggested later.

Figure \ref{fig:gapdepth2} shows how $\delta(q)$ depends on $\mathcal{M}$ and $\alpha$.  The break point $q_{\rm NL}$ very simply scales as 

\begin{equation}
    q_{\rm NL} = 1.04 \mathcal{M}^{-3},
\end{equation}
with no dependence on $\alpha$.

This is consistent with the picture that the break point occurs when the spiral wave becomes strongly nonlinear; $q \sim \mathcal{M}^{-3}$ is also the correct scaling for the thermal mass, or the mass ratio at which the Hill radius becomes larger than the disk scale height.

Apparently this non-linearity causes the torque deposited by shocks to be somewhat weaker than would be predicted by analytical theory.  This was also noted by \cite{2017PASJ...69...97K}, who suggested a nonlinear adjustment to the \cite{2013ApJ...769...41D} scaling by a factor of $0.4$ (referred to as $f_{\rm NL}$ in the Kanagawa paper).  However, it was not clear at what point their adjustment should be applied.  

Additionally, even with this adjustment it is clear that the scaling will be incorrect as the gap deepens; the $q^3$ term in $\delta(q)$ might be due to the fact that the wave is no longer being predominantly excited at the bottom of the gap, but at the higher-density gap walls \citep{2018MNRAS.479.1986G}.  Regardless of the reason, this effect becomes important around $q \sim q_w$, whose scaling we can also determine empirically.

\cite{2018MNRAS.479.1986G} predicted that the gap depth scaling should change due to excitation of torque at the gap walls.  This transition was predicted to occur when

\begin{equation}
    q_w \sim \alpha^{1/5} / \mathcal{M}^{14/5},
\end{equation}
or, in terms of $q_{\rm NL}$,

\begin{equation}
    q_w \sim q_{\rm NL} (\alpha \mathcal{M})^{1/5}.
\end{equation}

In practice, instead of a $1/5$ power, a steeper scaling is found:
\begin{equation}
    q_w = 34~ q_{\rm NL} (\alpha \mathcal{M})^{1/2}.
\end{equation}
Interestingly, this value of $q_w$ is the mass ratio at which the gap attains the depth $\Sigma_{\rm gap} \approx 0.02 \Sigma_0$ according to the \cite{2013ApJ...769...41D} formula.  So it makes some sense that excitation at the gap walls would dominate when the gap becomes sufficiently deep.

Additionally, \cite{2018MNRAS.479.1986G} predicted the scaling $\delta(q) \sim (q/q_w)^2$ for this regime, which is shallower than the scaling observed here, hence the power of three in Equation (\ref{eqn:delta}).

The present study only investigated disks for which $q_{\rm NL} < q_w$.  That is, for $\alpha > 8.7 \times 10^{-4}/\mathcal{M}$.  For sufficiently low viscosity, excitation of torque at the gap walls becomes important before nonlinearity in the wave affects the scaling.  It is possible that the \cite{2018MNRAS.479.1986G} scalings become valid in this low-viscosity regime, as that is the regime in which they were calculated.

Finally, with this formula for $\delta(q)$, it is now possible to plot $\Sigma_{\rm gap}(q,\alpha,\mathcal{M})$, comparing numerical results to the new analytical formula, for all mass ratios and disk models considered.  This is shown in Figure \ref{fig:gapdepth3}.  This figure also shows how this scaling breaks down when the gap becomes too deep and unstable.  This regime will not be characterized in this study, except to note that it appears possible to describe the unstable regime as hitting a floor in the density of 

\begin{equation}
    \Sigma_{\rm gap} = 0.015 \Sigma_0 e^{-600 q},
\end{equation}
indicated by the red curve in the figure.  This ``floor" did not work for all models (curiously, the lowest viscosity model in the lower panel achieved a much deeper gap) but it appears to work for many disks.

\begin{figure}
\includegraphics[width=3.4 in]{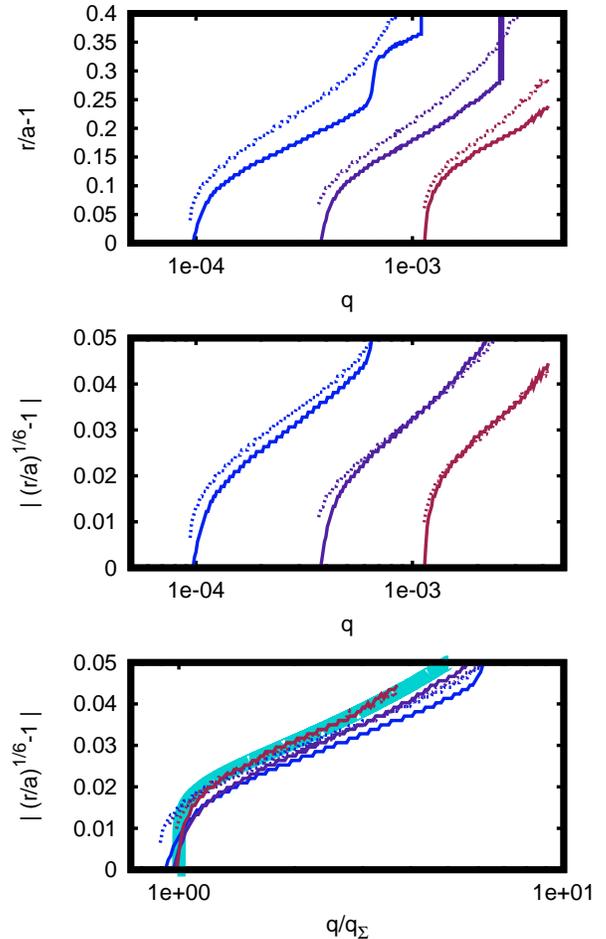}
\caption{Top Panel:  Gap radii $R_\pm$ are plotted at different gap depths, i.e. the maximum and minimum radius which attain the surface density $\Sigma_{\rm gap}/\Sigma_0 = 0.9$, $0.5$, and $0.1$.  Center Panel: While $R_+$ and $R_-$ are different distances from the planetary radius $a$, it turns out that $R_+^{1/6}$ is about the same distance from $a^{1/6}$ as $R_-^{1/6}$.  Note that all scalings at different depths all appear to have a similar shape.  Bottom Panel:  A single formula is found to fit all six curves by taking the center panel and dividing the x-axis by $q_{\Sigma}$, the mass ratio at which the gap is depleted to that depth.  In this figure, the curves are found to fit the functional form $((x^3-1)/D^3)^{1/6}$, for $D = 2000$.  The choice of $D$ was fit to the curves of $R_{\pm}^{0.1}$. }
\label{fig:gapwidth1}
\end{figure}

\subsection{Gap Width}

\begin{figure}
\includegraphics[width=3.4 in]{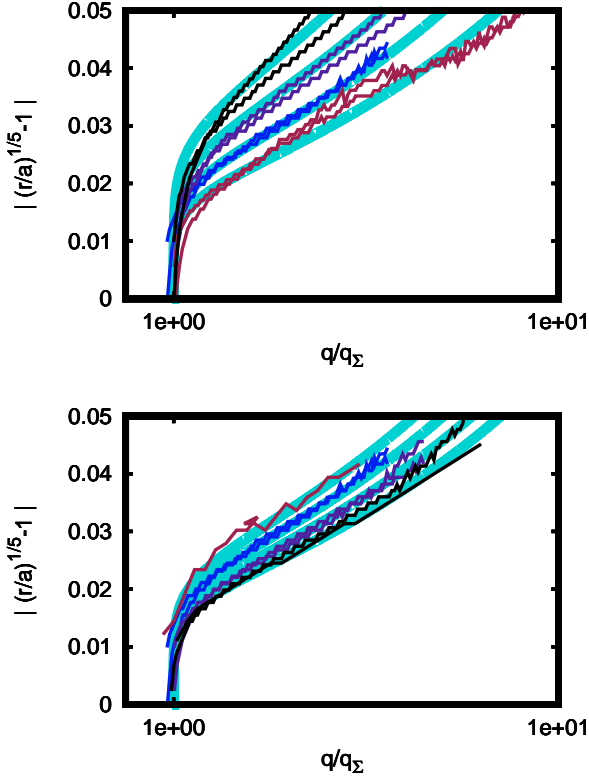}
\caption{Scaling of $D$ with $\mathcal{M}$ is determined by comparing the gap formula (\ref{eqn:width}) with the inner and outer radii where $\Sigma = 0.1 \Sigma_0$ ($R_{\pm}^{0.1}$) for the different disk models considered.}
\label{fig:gapwidth2}
\end{figure}


Gap width in this study is measured at three different surface densities: $\Sigma / \Sigma_0 = 0.1$, $0.5$, and $0.9$.  The inner and outer radii with this surface density are measured and plotted in the top panel of Figure \ref{fig:gapwidth1} as a function of mass ratio for a particular disk model $\mathcal{M} = 20$, $\alpha = 0.01$.

The gap is not symmetric; $R_{+}$ and $R_{-}$ are not the same distance from the planet.  However, through experimentation it was found that $(R_{+})^{1/6}$ and $(R_{-})^{1/6}$ are roughly equidistant from $a^{1/6}$.  The middle panel of Figure \ref{fig:gapwidth1} shows this, by plotting $| (R_{\pm}/a)^{1/6} - 1 |$ for a range of planet masses and for the three different gap depths measured.

After viewing the central panel of Figure \ref{fig:gapwidth1}, it becomes clear that the three curves corresponding to gap widths at different depths appear to obey a very similar scaling to one another, and that they should approximately lie on top of one another if one re-scales the mass ratio by $q_{\Sigma}$, the mass ratio at which the gap attains the depth $\Sigma$.

This final curve is fit by a power-law.  For $\mathcal{M} = 20$ and $\alpha = 0.01$, this curve is found to be well-fit by the function

\begin{equation}
    ( (r/a)^{1/6} - 1 )^6 = \frac{(q/q_{\Sigma})^3 - 1}{D^3},
    \label{eqn:width}
\end{equation}

where $D = 2000$ for this disk.

This provides an (apparently) universal scaling for the gap width with mass ratio.  However, we have yet to establish how the quantity $D$ varies with $\mathcal{M}$ and $\alpha$.

In practice, we find this by fitting the same curves for different disk models (see Figure \ref{fig:gapwidth2}).  The scaling found was

\begin{equation}
    D(\mathcal{M},\alpha) = 7~ \mathcal{M}^{3/2} / \alpha^{1/4}.
    \label{eqn:D}
\end{equation}

One can compare this to the measured scaling found by \cite{2015ApJ...806L..15K}.  Note, in the regime where the width $\Delta \ll a$ but $q \gg q_{\Sigma}$, equation (\ref{eqn:width}) becomes:

\begin{equation}
    (\Delta/a)^6 \sim \left( \frac{q}{q_{\Sigma} D} \right)^3.
    \label{eqn:wscale}
\end{equation}

Now, apply the \cite{2012ApJ...755....7D} scaling for $q_{\Sigma}$, i.e. the scaling of $q$ with $\alpha$ and $\mathcal{M}$ to fix a specific gap depth $\Sigma$:

\begin{equation}
    q_{\Sigma}^2 \sim \alpha/\mathcal{M}^5
\end{equation}
and therefore,

\begin{equation}
    \Delta/a \sim \frac{ \sqrt{q} ( \mathcal{M}^5 / \alpha )^{1/4} }{ D^{1/2} }.
\end{equation}

Using the measured scaling for $D$, one arrives at:

\begin{equation}
    \Delta/a \sim \sqrt{q \mathcal{M}} / \alpha^{1/8}.
\end{equation}

Comparing this to the \cite{2015ApJ...806L..15K} scaling:

\begin{equation}
    \Delta/a \sim \sqrt{q}~ \mathcal{M}^{3/4} / \alpha^{1/4}.
\end{equation}

The scalings are reasonably close, off by a factor of $(\alpha/\mathcal{M}^2)^{1/8}$.  Any discrepancies could potentially be explained by the fact that the gap width does not really follow a power-law.

Figure \ref{fig:gapwidth2} shows that our scaling is close to the numerical curves $R_{\pm}(q,\alpha,\mathcal{M})$ for all disk models considered.  It is worth noting that this scaling fails once the disk becomes unstable and turbulent; the gap widens significantly in that case.  It is possible that this is related to the different scalings found by \cite{2015ApJ...806L..15K}.  A future study can address this discrepancy.

\section{Comparing the Model to Steady-State Disks}
\label{sec:compare}

\begin{figure}
\includegraphics[width=3.4 in]{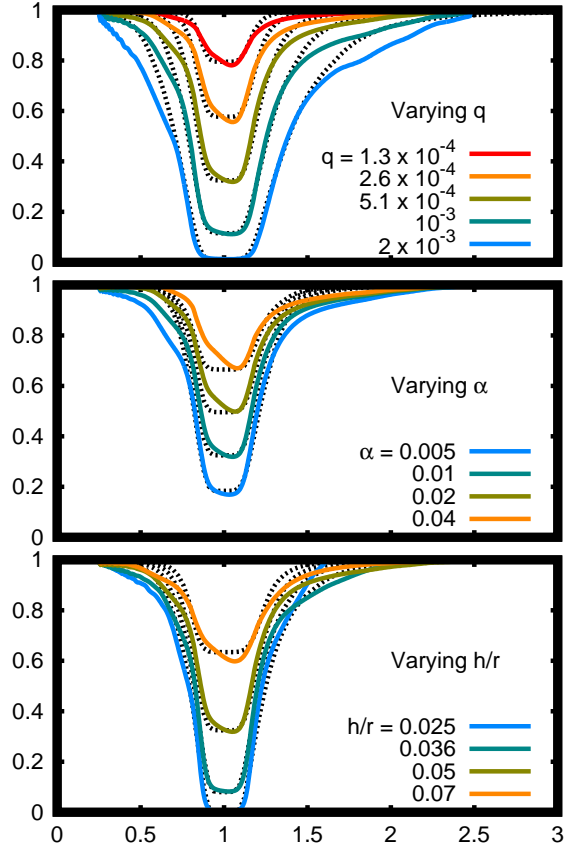}
\caption{Comparison between the gap model and a wide range of gaps numerically calculated in a previous study of planet disk interactions \citep{2015ApJ...806..182D}.  Black dashed curves are the analytical model, and colored solid curves are the numerically calculated gap profiles in steady-state disks.  The fiducial disk model used $q = 5.1 \times 10^{-4}$, $\alpha = 0.01$, $h/r = 0.05$ (or $\mathcal{M} = 20$).  Each panel picks one of these parameters and varies it, keeping the others fixed at their fiducial values.  For all planet masses, disk aspect ratios and viscosities considered, the analytical formula is a good model for the gap structure.  The two most major points where the model breaks down are (1) when the planet triggers instabilities in the disk that lead to vortex production, and (2) deviations further from the planet, around $r/a = 0.5$ and $1.5$, especially in the top panel.  This ``second step" feature was predicted by \cite{2018MNRAS.479.1986G}, but is not accounted for in this study.}
\label{fig:compare}
\end{figure}

Equipped with the scaling (\ref{eqn:width}), along with the formula for gap depth (\ref{eqn:depth}), it becomes possible to build a complete formula for $\Sigma(r)$.

First, note that ``$q_{\Sigma}$" is just the inverse function of $\Sigma_{\rm gap}(q)$.  Therefore, one can simply write

\begin{equation}
    \Sigma(r) = \Sigma_{\rm gap}(\tilde q),
\end{equation}
where $\tilde q$ is given by solving equation (\ref{eqn:width}) for $q_{\Sigma}$:

\begin{equation}
    \tilde q(r) \equiv \frac{q}{ (1 + D^3 ( (r/a)^{1/6} - 1 )^6 )^{1/3} }.
    \label{eqn:final1}
\end{equation}

Then,

\begin{equation}
    \Sigma(r) = \frac{\Sigma_0}{1 + \frac{0.45}{3 \pi} \frac{\tilde q(r)^2 \mathcal{M}^5}{\alpha} \delta(\tilde q(r)) }.
    \label{eqn:final2}
\end{equation}

Equations (\ref{eqn:final1}) and (\ref{eqn:final2}), along with equation (\ref{eqn:delta}) for $\delta(q)$ and equation (\ref{eqn:D}) for $D$, provide a complete formula for $\Sigma(r)$, given any $q$, $\alpha$, and $\mathcal{M}$.

We can put this model to the test by comparing with steady-state numerical calculations of gap-opening planets.  For this, we use steady-state models generated from the disk-planet study of \cite{2015ApJ...806..182D} (Figure \ref{fig:compare}).  Excellent agreement is found between the numerical and analytical gap profiles for a wide range of gap-opening planets.  In particular, the model is able to capture both shallow and deep gaps, whereas the analytical model of \cite{2015ApJ...807L..11D} was only able to capture shallow gaps.

\begin{figure}
\includegraphics[width=3.4 in]{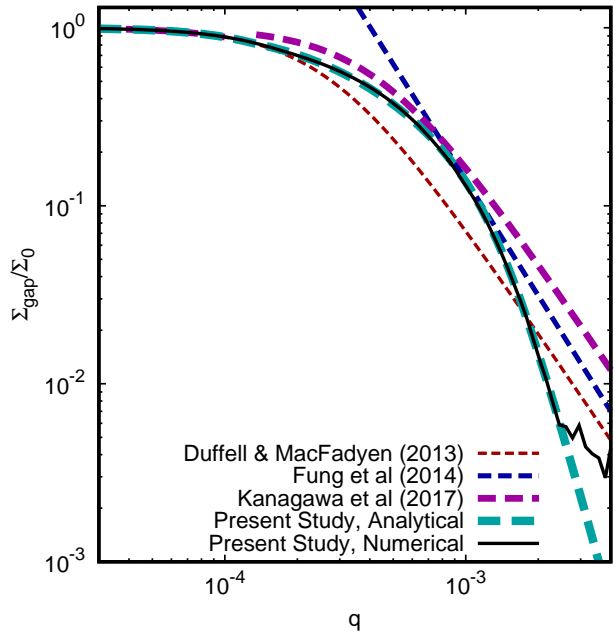}
\caption{Comparison between the numerical calculations (black solid curve) with analytical scalings derived in the present work and various analytical scalings in previous works.  The power-law scalings calculated by \cite{2014ApJ...782...88F} appear to recover the tangent line at Jupiter's mass ($q = 10^{-3}$), which is the vicinity of parameter space they were measured in that study.  Above Jupiter's mass, the scaling becomes much more steep, $\propto q^{-5}$.  \cite{2018MNRAS.479.1986G} predicted this steepening, though they predicted a scaling of $q^{-4}$ in this regime.}
\label{fig:anly_comp}
\end{figure}

Figure \ref{fig:anly_comp} shows how the new formula for gap depths compares with other formulas in the literature.  By design, the model agrees with \cite{2013ApJ...769...41D} in the low-mass regime, but deviates in the high-mass regime (as also shown in Figure \ref{fig:gapdepth1}).  The scaling measured empirically by \cite{2014ApJ...782...88F} appears to precisely recover the tangent line near Jupiter's mass.  The scaling of \cite{2017PASJ...69...97K} with the nonlinear correction $f_{\rm NL} = 0.4$ reasonably captures gap depths up to Jupiter's mass, but fails above this mass.  The scaling here is the only one to recover the gap depth in all regimes, up to a few times Jupiter's mass (after which point instabilities in the disk may take over, depending on the disk properties).

\section{Discussion} \label{sec:disc}

The calculations presented here can be used to model gaps in resolved images of protoplanetary disks.  The formula for $\Sigma(r)$ presented here, while being largely empirical in nature, may help guide analytical studies to come up with an explanation for these formulas.  Understanding the various break points in $\delta(q)$, for example, will require additional analytical work, which can be tested against this empirical formula rather than running full numerical hydrodynamics calculations.

In order to make these scalings as easy-to-use as possible, this 1D gap model has been built into a short, simple code to rapidly generate synthetic disk profiles.  This code is available for download at \texttt{https://github.com/duffell/Duffell\_Gap}.  The code is written in both C and Python and is freely available under the MIT license.

The scalings found here break down when the density gradients become steep enough to trigger instabilities in the disk, leading to the production of vortices.  It was found that the onset of instability was suppressed by extending the boundary further from the planet and increasing the planet mass slowly.  More care must be taken to determine whether the numerical calculation is faithfully capturing the instability.  Future work will require a meticulous survey to characterize the unstable regime.  The analytical model is accurate enough in the stable regime that it may be possible to use it to analytically calculate the point where the disk should become unstable to different instabilities such as the Rayleigh instability or Rossby wave instability.  This idea is left for a future study.

Another obvious deviation between the empirical model and the numerical results is the ``tails" in the gap located many scale heights away from the planet (see for example the top panel of Figure \ref{fig:compare}, at $r = 0.5 a$ and $r = 1.5 a$).  These deviations may be reasonable to model empirically as well \citep[in fact, a very similar feature was predicted by][]{2018MNRAS.479.1986G}, but modeling this structure was not attempted in this study because of a lack of confidence in the numerical results for these ``tails"; they may be affected by the inner and outer boundary.  This detail may be resolvable in a future study.

Finally, a very clear feature that occurs for many of the shallow gaps in Figure \ref{fig:compare} is an asymmetry in the gap; the outer side of the gap is consistently more depleted than the inner side.  The likely cause of this is the same torque imbalance that leads to Type I migration; the outer Lindblad resonances consistently excite stronger torques than the inner Lindblad resonances.  This could potentially used as an indicator to observers as to whether the gap is caused by a planet; if a gap were observed whose asymmetry went in the opposite direction, it might be difficult to model the gap using a planet.  However, modeling this asymmetry was not attempted in this study; it appears to occur mainly for shallow gaps where $\Sigma_{\rm gap} \sim 0.5 \Sigma_0$.

\acknowledgments

I am grateful to Karin \"Oberg, Jane Huang, Sivan Ginzburg, Andrew MacFadyen, Daniel D'Orazio, Zoltan Haiman and Andrea Derdzinski for helpful comments and discussions.

\bibliographystyle{apj} 
\bibliography{diskbib}

\begin{thebibliography}{}
\expandafter\ifx\csname natexlab\endcsname\relax\def\natexlab#1{#1}\fi

\bibitem[{{Crida} {et~al.}(2006){Crida}, {Morbidelli}, \&
  {Masset}}]{2006Icar..181..587C}
{Crida}, A., {Morbidelli}, A., \& {Masset}, F. 2006, \icarus, 181, 587

\bibitem[{{Duffell}(2015{\natexlab{a}})}]{2015ApJ...807L..11D}
{Duffell}, P.~C. 2015{\natexlab{a}}, \apjl, 807, L11

\bibitem[{{Duffell}(2015{\natexlab{b}})}]{2015ApJ...806..182D}
---. 2015{\natexlab{b}}, \apj, 806, 182

\bibitem[{{Duffell}(2016)}]{2016ApJS..226....2D}
---. 2016, \apjs, 226, 2

\bibitem[{{Duffell} \& {MacFadyen}(2012)}]{2012ApJ...755....7D}
{Duffell}, P.~C., \& {MacFadyen}, A.~I. 2012, \apj, 755, 7

\bibitem[{{Duffell} \& {MacFadyen}(2013)}]{2013ApJ...769...41D}
---. 2013, \apj, 769, 41

\bibitem[{{Fung} {et~al.}(2014){Fung}, {Shi}, \&
  {Chiang}}]{2014ApJ...782...88F}
{Fung}, J., {Shi}, J.-M., \& {Chiang}, E. 2014, \apj, 782, 88

\bibitem[{{Ginzburg} \& {Sari}(2018)}]{2018MNRAS.479.1986G}
{Ginzburg}, S., \& {Sari}, R. 2018, \mnras, 479, 1986

\bibitem[{{Goodman} \& {Rafikov}(2001)}]{2001ApJ...552..793G}
{Goodman}, J., \& {Rafikov}, R.~R. 2001, \apj, 552, 793

\bibitem[{{Guzm{\'a}n} {et~al.}(2018){Guzm{\'a}n}, {Huang}, {Andrews},
  {Isella}, {P{\'e}rez}, {Carpenter}, {Dullemond}, {Ricci}, {Birnstiel},
  {Zhang}, {Zhu}, {Bai}, {Benisty}, {{\"O}berg}, \&
  {Wilner}}]{2018ApJ...869L..48G}
{Guzm{\'a}n}, V.~V., {Huang}, J., {Andrews}, S.~M., {et~al.} 2018, \apjl, 869,
  L48

\bibitem[{{Hammer} {et~al.}(2017){Hammer}, {Kratter}, \&
  {Lin}}]{2017MNRAS.466.3533H}
{Hammer}, M., {Kratter}, K.~M., \& {Lin}, M.-K. 2017, \mnras, 466, 3533

\bibitem[{{Huang} {et~al.}(2018{\natexlab{a}}){Huang}, {Andrews}, {Cleeves},
  {{\"O}berg}, {Wilner}, {Bai}, {Birnstiel}, {Carpenter}, {Hughes}, {Isella},
  {P{\'e}rez}, {Ricci}, \& {Zhu}}]{2018ApJ...852..122H}
{Huang}, J., {Andrews}, S.~M., {Cleeves}, L.~I., {et~al.} 2018{\natexlab{a}},
  \apj, 852, 122

\bibitem[{{Huang} {et~al.}(2018{\natexlab{b}}){Huang}, {Andrews}, {Dullemond},
  {Isella}, {P{\'e}rez}, {Guzm{\'a}n}, {{\"O}berg}, {Zhu}, {Zhang}, {Bai},
  {Benisty}, {Birnstiel}, {Carpenter}, {Hughes}, {Ricci}, {Weaver}, \&
  {Wilner}}]{2018ApJ...869L..42H}
{Huang}, J., {Andrews}, S.~M., {Dullemond}, C.~P., {et~al.} 2018{\natexlab{b}},
  \apjl, 869, L42

\bibitem[{{Huang} {et~al.}(2018{\natexlab{c}}){Huang}, {Andrews}, {P{\'e}rez},
  {Zhu}, {Dullemond}, {Isella}, {Benisty}, {Bai}, {Birnstiel}, {Carpenter},
  {Guzm{\'a}n}, {Hughes}, {{\"O}berg}, {Ricci}, {Wilner}, \&
  {Zhang}}]{2018ApJ...869L..43H}
{Huang}, J., {Andrews}, S.~M., {P{\'e}rez}, L.~M., {et~al.} 2018{\natexlab{c}},
  \apjl, 869, L43

\bibitem[{{Kanagawa} {et~al.}(2015{\natexlab{a}}){Kanagawa}, {Muto}, {Tanaka},
  {Tanigawa}, {Takeuchi}, {Tsukagoshi}, \& {Momose}}]{2015ApJ...806L..15K}
{Kanagawa}, K.~D., {Muto}, T., {Tanaka}, H., {et~al.} 2015{\natexlab{a}},
  \apjl, 806, L15

\bibitem[{{Kanagawa} {et~al.}(2017){Kanagawa}, {Tanaka}, {Muto}, \&
  {Tanigawa}}]{2017PASJ...69...97K}
{Kanagawa}, K.~D., {Tanaka}, H., {Muto}, T., \& {Tanigawa}, T. 2017, \pasj, 69,
  97

\bibitem[{{Kanagawa} {et~al.}(2015{\natexlab{b}}){Kanagawa}, {Tanaka}, {Muto},
  {Tanigawa}, \& {Takeuchi}}]{2015MNRAS.448..994K}
{Kanagawa}, K.~D., {Tanaka}, H., {Muto}, T., {Tanigawa}, T., \& {Takeuchi}, T.
  2015{\natexlab{b}}, \mnras, 448, 994

\bibitem[{{Kanagawa} {et~al.}(2018){Kanagawa}, {Tanaka}, \&
  {Szuszkiewicz}}]{2018ApJ...861..140K}
{Kanagawa}, K.~D., {Tanaka}, H., \& {Szuszkiewicz}, E. 2018, \apj, 861, 140

\bibitem[{{P{\'e}rez} {et~al.}(2018){P{\'e}rez}, {Benisty}, {Andrews},
  {Isella}, {Dullemond}, {Huang}, {Kurtovic}, {Guzm{\'a}n}, {Zhu}, {Birnstiel},
  {Zhang}, {Carpenter}, {Wilner}, {Ricci}, {Bai}, {Weaver}, \&
  {{\"O}berg}}]{2018ApJ...869L..50P}
{P{\'e}rez}, L.~M., {Benisty}, M., {Andrews}, S.~M., {et~al.} 2018, \apjl, 869,
  L50

\bibitem[{{Rafikov}(2002)}]{2002ApJ...569..997R}
{Rafikov}, R.~R. 2002, \apj, 569, 997

\bibitem[{{Varni{\`e}re} {et~al.}(2004){Varni{\`e}re}, {Quillen}, \&
  {Frank}}]{2004ApJ...612.1152V}
{Varni{\`e}re}, P., {Quillen}, A.~C., \& {Frank}, A. 2004, \apj, 612, 1152

\bibitem[{{Ward}(1997)}]{1997Icar..126..261W}
{Ward}, W.~R. 1997, \icarus, 126, 261

\end{thebibliography}

\end{document}